\documentclass[preprint,3p,fleqn,sort]{elsarticle}
\usepackage[labelfont=bf, singlelinecheck=false,
            justification=justified]{caption}
\usepackage{xcolor}
\usepackage{amsmath,amssymb}
\usepackage{subcaption}
\usepackage{mathrsfs}
\usepackage{stmaryrd}
\usepackage{booktabs}
\usepackage{microtype}
\usepackage{mathtools}
\usepackage{multirow}
\usepackage{enumitem}
\usepackage{setspace}
\usepackage{placeins}
\usepackage{etoolbox,siunitx}
\usepackage{bm}
\robustify\bfseries

\makeatletter
\g@addto@macro\normalsize{%
	\setlength\abovedisplayskip{5pt}
	\setlength\belowdisplayskip{5pt}
	\setlength\abovedisplayshortskip{6pt}
	\setlength\belowdisplayshortskip{6pt}
}
\makeatother

\usepackage[T1]{fontenc}

\SetSymbolFont{stmry}{bold}{U}{stmry}{m}{n}

\let\rho\varrho

\usepackage{accsupp}
\newcommand*{\llbrace}{%
  \BeginAccSupp{method=hex,unicode,ActualText=2983}%
    \textnormal{\usefont{OMS}{lmr}{m}{n}\char102}%
    \hspace*{-3pt}
    \textnormal{\usefont{OMS}{lmr}{m}{n}\char102}%
  \EndAccSupp{}%
}
\newcommand*{\rrbrace}{%
  \BeginAccSupp{method=hex,unicode,ActualText=2984}%
    \textnormal{\usefont{OMS}{lmr}{m}{n}\char103}%
        \hspace*{-3pt}
    \textnormal{\usefont{OMS}{lmr}{m}{n}\char103}%
  \EndAccSupp{}%
}

\usepackage{lineno,hyperref}
\modulolinenumbers[5]

\renewcommand{\vec}[1]{\ensuremath{\boldsymbol #1}}

\newcommand{\pderivative}[2]{\frac{\partial #1}{\partial #2}}
\newcommand{\eg}{\textit{e.g.}~}
\newcommand{\ie}{\textit{i.e.}~}
\renewcommand{\vec}[1]{\ensuremath{\boldsymbol #1}}
\newcommand{\He}{\ensuremath{\bm{\mathcal{H}}}}
\renewcommand{\H}{\ensuremath{\bm{\hat{\mathcal{H}}}}}
\newcommand{\mat}[1]{\ensuremath{\mathbf{#1}}}
\newcommand{\avg}[1]{\ensuremath{\llbrace#1\rrbrace}}
\newcommand{\avgln}[1]{\ensuremath{#1^\mathrm{ln}}}

\newcommand{\jump}[1]{\ensuremath{\left\llbracket #1 \right\rrbracket}}
\newcommand{\rholn}{\ensuremath{\rho\textsuperscript{ln}}}
\newcommand{\betaln}{\ensuremath{\beta\textsuperscript{ln}}}
\newcommand{\pln}{\ensuremath{p\textsuperscript{ln}}}
\newcommand{\pavg}{\ensuremath{\avg{p}}}
\newcommand{\uavg}{\overline{\lVert \vec{u} \rVert^2}}
\newcommand{\betaavg}{\ensuremath{\overline{\beta^2}}}

\newcommand{\Eline}{\ensuremath{\overline{E}}}

\begin{document}

\begin{frontmatter}

\title{A Novel Averaging Technique for Discrete Entropy-Stable Dissipation Operators for Ideal MHD}

\author[physik]{Dominik Derigs\corref{mycorrespondingauthor}}
\cortext[mycorrespondingauthor]{Corresponding author}
\ead{derigs@ph1.uni-koeln.de}
\author[mathematik]{Andrew R.~Winters}
\author[mathematik]{Gregor J.~Gassner}
\author[physik]{Stefanie Walch}
\address[physik]{I.\,Physikalisches Institut, Universit\"at zu K\"oln, Z\"ulpicher Stra\ss{}e~77, 50937 K\"oln}
\address[mathematik]{Mathematisches Institut, Universit\"at zu K\"oln, Weyertal 86-90, 50931 K\"oln}

\numberwithin{equation}{section}


\begin{abstract}
	Entropy stable schemes can be constructed with a specific choice of the numerical flux function. First, an entropy conserving flux is constructed. Secondly, an entropy stable dissipation term is added to this flux to guarantee dissipation of the discrete entropy. 
	Present works in the field of entropy stable numerical schemes are concerned with thorough derivations of entropy conservative fluxes for ideal MHD.
	However, as we show in this work, if the dissipation operator is not constructed in a very specific way, it cannot lead to a generally stable numerical scheme.

	The two main findings presented in this paper are that the entropy conserving flux of Ismail \& Roe can easily break down for certain initial conditions commonly found in astrophysical simulations, and that special care must be taken in the derivation of a discrete dissipation matrix for an entropy stable numerical scheme to be robust.
	
	We present a convenient novel averaging procedure to evaluate the entropy Jacobians of the ideal MHD and the compressible Euler equations that yields a discretization with favorable robustness properties.
\end{abstract}

\end{frontmatter}

\section{Introduction}
The applications of ideal magnetohydrodynamics (MHD) are ubiquitous in science and engineering. Accordingly, the design of numerical schemes for the approximation of this particular set of hyperbolic conservation laws has undergone extensive development.
The ideal MHD model assumes that a fluid is a good electric conductor and neglects non-ideal effects, \eg viscosity or resistivity. It is governed by a system of conservation laws together with the divergence-free condition
\begin{align}\label{eq:3DIDEALMHD}
\pderivative{}{t}\, \vec{q} + \nabla\cdot\vec{f} =
\pderivative{}{t}
\begin{bmatrix}
\rho \\[1mm] \rho\vec{u} \\[1mm] E \\[1mm] \vec{B} \end{bmatrix}
+
&\nabla\cdot
\begin{bmatrix}
\rho\vec{u} \\[1mm]
\rho(\vec{u}\otimes\vec{u}) + \left(p+\frac{1}{2}\|\vec{B}\|^2\right)\mat{I}-\vec{B}\otimes\vec{B} \\[1mm]
\vec{u}\left(\frac{1}{2}\rho\|\vec{u}\|^2 + \frac{\gamma p}{\gamma - 1} + \|\vec{B}\|^2 \right) - \vec{B}(\vec{u}\cdot\vec{B}) \\[1mm]
\vec{u}\otimes\vec{B} - \vec{B}\otimes\vec{u}
\end{bmatrix} = \vec{0}, 
\end{align}
where $\rho$, $\rho\vec{u}$, and $E$ are the mass, momenta, and total specific energy of the plasma system, $p$ is the thermal pressure, $\mat{I}$ is the $3 \times 3$ identity matrix, and $\vec{B}$ is the magnetic field, also referred to as magnetic flux density. $\vec{f}$ is the multidimensional flux function.

It is well-known that solutions to \eqref{eq:3DIDEALMHD} may contain discontinuities in the form of shocks, even for smooth initial data. Hence, solutions are sought in the weak sense \cite{Fjordholm2012}. However, weak solutions are not unique and need to be supplemented with extra admissibility criteria.
Following the work of \eg \cite{Derigs2016,Winters2016,Fjordholm2011,Tadmor1987}, we use the concept of entropy stability to construct discretizations that agree with the second law of thermodynamics.

In this paper we describe a technique suitable for the derivation of a flux dissipation term that guarantees entropy stability.
Sec.~\ref{scn:NumDiffOp} provides the necessary background of entropy numerical fluxes. In Sec.~\ref{scn:ECfailure} we motivate our choice for the baseline entropy conserving flux and apply our technique to derive a simple dissipation operator in Sec.~\ref{scn:Derivation}. In Sec.~\ref{scn:comp} we investigate the computational costs of our modifications. Finally, in Sec.~\ref{scn:numexp}, we revisit the two main findings of this work using simple numerical tests from the field of astrophysics.

\section{Entropy stable numerical flux functions}\label{scn:NumDiffOp}
For smooth solutions, one can design numerical methods to be \textbf{\emph{entropy conservative}} if, discretely, the local changes of entropy are the same as predicted by the continuous entropy conservation law. For discontinuous solutions, the approximation is said to be \textbf{\emph{entropy stable}} if the entropy always possesses the correct sign and the numerical scheme produces more entropy than an entropy conservative scheme and satisfies the entropy inequality (where we use the mathematical notation that entropy is a decaying function)
\begin{equation}\label{eq:EntropyInequality}
\pderivative{}{t}S + \nabla \cdot (\vec{u} S) \le 0.
\end{equation}
with the entropy density $S = -\frac{\rho s}{\gamma - 1}$, the corresponding entropy flux $\vec{u} S$, and the specific nondimensional thermodynamic entropy $s = \ln\big(p \rho^{-\gamma}\big)$, where $\gamma = \frac{c_p}{c_v}$ is the ratio of specific heats \cite{Derigs2016}.
Because entropy conservative schemes will produce high-frequency oscillations near shocks (see \eg \cite{Winters2016}), we need to add a carefully designed dissipation term to ensure that the entropy is guaranteed to dissipate.

Therefore, in order to create an entropy stable numerical approximation, we use a suitable entropy conserving flux as a base and add a numerical dissipation term. The resulting numerical flux is of the form
\begin{equation}\label{eq:entropystableflux}
	\vec{f}^* = \vec{f}^{*,ec} - \frac{1}{2} \mat{D} \jump{\vec{q}},
\end{equation}
where $\mat{D}$ is a suitable dissipation operator, and $\vec{q}$ is the vector of conserved quantities. We define the jump operator as $\jump{\cdot} = (\cdot)_{\rm R} - (\cdot)_{\rm L}$. Of utmost concern for entropy stability is to formulate the dissipation term in \eqref{eq:entropystableflux} such that the numerical flux fulfills the entropy inequality \eqref{eq:EntropyInequality}.

If we make the choice of $\mat{D}$ to be
\begin{equation}
	\mat{D} = |\lambda_{\rm max}| \mat{I},
\end{equation}
where $\lambda_{\rm max}$ is the largest eigenvalue of the ideal MHD system, we can rewrite the dissipation term
\begin{subequations}
\begin{align}
\frac{1}{2} \mat{D} \jump{\vec{q}} &= \frac{1}{2}  |\lambda_{\rm max}| \mat{I} \jump{\vec{q}}, \label{eq:dissipation_term1}\\
 &= \frac{1}{2} |\lambda_{\rm max}| \He{} \jump{\vec{v}}, \label{eq:dissipation_term2}
\end{align}
\end{subequations}
where $\vec{v}$ is the vector of entropy variables and $\He{} = \pderivative{\vec{v}}{\vec{q}}$ is a matrix that relates the variables in conserved and entropy space. This particularly simple choice for the dissipation term leads to a \emph{scalar dissipation term}.
Note that a scalar dissipation term cannot resolve contact discontinuities exactly, as it will always add dissipation on surfaces that separate zones of different densities.
The reformulation of the dissipation term to incorporate the jump in entropy variables (rather than the jump in conservative variables) is done to be able to guarantee entropy stability \cite{Barth1999}.
The question is how should the entropy Jacobian be evaluated at the interface, where values from left and right are available.

\section{Break down of Ismail and Roe's entropy conservative scheme and the KEPEC flux}\label{scn:ECfailure}
First, we consider the widely used Ismail and Roe (IR) entropy conservative flux for Euler flows (see \eg \cite{Roe2006,IsmailRoe2009}).
Define the arithmetic mean and the logarithmic mean of any strictly positive quantity as
\begin{equation*}
	\avg{\cdot} = \frac{(\cdot)_{\rm L} + (\cdot)_{\rm R}}{2},\ %
	(\cdot)\textsuperscript{ln} = \frac{\jump{\cdot}}{\jump{\ln(\cdot)}},
\end{equation*}
where a stable numerical algorithm to compute the logarithmic mean when $(\cdot)_{\rm R} \approx (\cdot)_{\rm L}$ is given in \cite[App.~B]{IsmailRoe2009}.

Using the parameter vector $\vec{z}$, the mass flux at an interface is given by
\begin{equation}\label{eq:IRMass}
f_{\rho}^{\rm IR} = \tilde{\rho}\tilde{u}, \quad \mbox{where}\quad \tilde{\rho} = \avg{z_1}\avgln{z_3}, \quad\mbox{and}\quad\tilde{u} = \frac{\avg{z_2}}{\avg{z_1}} \quad\mbox{with}\quad 
\vec{z} = \left[\sqrt{\frac{\rho}{p}},\sqrt{\frac{\rho}{p}}u,\sqrt{\rho p}\right]^\intercal .
\end{equation}
Consider the following initial conditions representing a very simplified form of a strong shock in a uniform moving medium,
\begin{equation}\label{eq:ICs}
\gamma=1.4,\quad \vec{p}_{\rm L} = \left[1, 10, 1\right],\quad\mbox{and}\quad \vec{p}_{\rm R} = \left[1, 10, 10^{-6}\right] \quad\text{with} \quad \vec{p} = \left[\rho, u, p\right].
\end{equation}
Using a Finite Volume (FV) update (see \eg \cite{Derigs2016}) with a CFL coefficient of $c=0.6$ we find
\begin{equation*}
	\rho'_{\rm L} \approx {\bm -}37.5 \quad\mbox{and}\quad\rho'_{\rm R} \approx 37.5
\end{equation*}
after the first time step (see also section~\ref{scn:numexp}).
Clearly, the greatly overestimated mass flux is unphysical and the scheme breaks down.
Note that the wrong mass flux cannot be compensated by the stabilization term, since any stabilization in the mass flux is proportional to the jump in density, which is zero according to \eqref{eq:ICs}.

If we, however, use the kinetic energy preserving entropy-conservative (KEPEC) flux presented by Chandrashekar \cite{Chandrashekar2012} and recently extended to ideal MHD by Winters \& Gassner \cite{Winters2016}, the mass flux is
\begin{equation}
f_{\rho}^{\rm KEPEC} = \avgln{\rho}\avg{u}.
\end{equation}
We obtain the updated densities
\begin{equation*}
	\rho'_{\rm L} \approx 1.0-\num{9e-10} \quad\mbox{and}\quad\rho'_{\rm R} \approx 1.0 + \num{9e-10}.
\end{equation*}
Note that even for much higher pressure jumps, there is no pathological behavior as seen in the mass flux with the $\vec{z}$ parametrization \eqref{eq:IRMass}.
Since example conditions given in \eqref{eq:ICs} are typical in astrophysical simulations, we conclude that the IR scheme is not suitable as an underlying entropy conservative scheme when constructing entropy stable schemes in a general way.
Therefore, we use the KEPEC flux as the baseline entropy conserving flux in \eqref{eq:entropystableflux}.

We further note that this example will equally fail in the case of ideal MHD.

\section{Derivation technique for the discrete entropy Jacobian}\label{scn:Derivation}
The entropy variables for an ideal gas with entropy $s=-(\gamma-1)\ln(\rho) - \ln(\beta)-\ln(2)$ are
\begin{equation}
\vec{v} = \pderivative{S}{\vec{q}} = \left[\frac{\gamma - s}{\gamma - 1}-\beta \lVert\vec{u}\rVert^2,\, 2\beta u, 2\beta v, 2\beta w, -2\beta, 2\beta B_1, 2\beta B_2, 2\beta B_3 \right]^\intercal,\quad \beta=\frac{\rho}{2p} \propto \frac{1}{T}.
\end{equation}

The goal is to derive the averages in such a way that the relation $\jump{\vec{q}} = \He{} \jump{\vec{v}}$ holds.
The entries of the matrix $\He{}$ are derived step-by-step through the solution of 64 linear equations:
\begin{equation}\label{eq:tobesolved}
\jump{\vec{q}} = \jump{\begin{bmatrix}\rho\\\rho u\\\rho v\\\rho w\\E\\B_1\\B_2\\B_3\end{bmatrix}} \stackrel{!}{=} \begin{bmatrix}
\He_{1,1} & \He_{1,2} & \dots & \dots & \He_{1,7} & \He_{1,8} \\
\He_{2,1} & \He_{2,2} & \dots & \dots & \He_{2,7} & \He_{2,8} \\
\vdots  & \vdots & \ddots & \ddots & \vdots & \vdots \\
\vdots  & \vdots & \ddots & \ddots & \vdots & \vdots \\
\He_{7,1} & \He_{7,2} & \dots & \dots & \He_{7,7} & \He_{7,8} \\
\He_{8,1} & \He_{8,2} & \dots & \dots & \He_{8,7} & \He_{8,8} \\
\end{bmatrix}
\jump{\begin{bmatrix}\frac{\gamma - s}{\gamma - 1}-\beta \lVert\vec{u}\rVert^2\\2\beta  u\\2\beta  v\\2\beta  w\\-2\beta \\2\beta B_1\\2\beta B_2\\2\beta B_3\end{bmatrix}}
 = \He{}\jump{\vec{v}}.
\end{equation}
The procedure is to multiply each row of $\He{}$ with the expanded jump in the entropy variables. By examining each equation individually, all unknown entries of the discrete matrix are found.

The derivation of the first row of $\He{}$ is straightforward and therefore serves as an excellent example for the derivation technique. First, we use properties of the linear jump operator \cite{Winters2016} to expand the jump in both the conservative and the entropy variables
\begin{align*}
&\jump{\vec{q}} = \resizebox{.885\hsize}{!}{$ %
	\jump{\begin{bmatrix}\rho \\ \rho u \\ \rho v \\ \rho w \\ E \\ B_1 \\ B_2 \\ B_3 \\\end{bmatrix}}
	=
	\begin{bmatrix}
	\jump{\rho} \\ \avg{\rho} \jump{u} + \avg{u} \jump{\rho} \\ \avg{\rho} \jump{v} + \avg{v} \jump{\rho}\\ \avg{\rho} \jump{w} + \avg{w} \jump{\rho} \\ \left(\frac{\avg{\beta^{-1}}}{2(\gamma-1)} + \frac{1}{2}\overline{\avg{\vec{u}^2}}\right)\jump{\rho}+\avg{\rho}\left(\avg{u}\jump{u} + \avg{v}\jump{v} + \avg{w}\jump{w}\right) - \frac{\avg{\rho}}{2\betaavg(\gamma-1)}\jump{\beta} + \sum\limits_{i=1}^{3}\avg{B_i}\jump{B_i} \\ \jump{B_1} \\ \jump{B_2} \\ \jump{B_3} \\
	\end{bmatrix}
$},\\
&\jump{\vec{v}} = \resizebox{.885\hsize}{!}{$ %
	\jump{\begin{bmatrix}\frac{\gamma - s}{\gamma - 1}-\beta \lVert\vec{u}\rVert^2\\2\beta  u\\2\beta  v\\2\beta  w\\-2\beta \\2\beta B_1\\2\beta B_2\\2\beta B_3\end{bmatrix}}
	=
	\begin{bmatrix}
	\frac{\jump{\rho}}{\rholn}+\frac{\jump{\beta}}{\betaln(\gamma-1)}-\Big(\avg{u^2}+\avg{v^2}+\avg{w^2}\Big)\jump{\beta}-2\avg{\beta}\Big( \avg{u}\jump{u} + \avg{v}\jump{v} + \avg{w}\jump{w} \Big) \\
	2 \avg{\beta}\jump{u} + 2 \avg{u}\jump{\beta} \\
	2 \avg{\beta}\jump{v} + 2 \avg{v}\jump{\beta} \\
	2 \avg{\beta}\jump{w} + 2 \avg{w}\jump{\beta} \\
	-2 \jump{\beta} \\
	2 \avg{\beta}\jump{B_1} + 2 \avg{B_1}\jump{\beta} \\
	2 \avg{\beta}\jump{B_2} + 2 \avg{B_2}\jump{\beta} \\
	2 \avg{\beta}\jump{B_3} + 2 \avg{B_3}\jump{\beta} \\
	\end{bmatrix}
	$}\ ,
\shortintertext{with}
	&{\small\betaavg = 2 \avg{\beta}^2 - \avg{\beta^2},\ \mbox{and}\ \overline{\avg{\vec{u}^2}} = \avg{u^2} + \avg{v^2} + \avg{w^2}.}
\end{align*}
According to \eqref{eq:tobesolved}, the entries of the first row of $\He{}$ can be obtained by solving
{\small
\begin{gather*}
\jump{\rho} = \He_{1,1}\left(\frac{\jump{\rho}}{\rholn} + \frac{\jump{\beta}}{\betaln (\gamma-1)} - \left(\avg{u^2} + \avg{v^2} + \avg{w^2}\right) \jump{\beta} - 2\avg{\beta}\Big(\avg{u}\jump{u} + \avg{v}\jump{v} + \avg{w}\jump{w}\Big)\right) \\
+\, \He_{1,2}\left( 2 \avg{\beta}\jump{u} + 2 \avg{u}\jump{\beta} \right) + \He_{1,3}\left( 2 \avg{\beta}\jump{v} + 2 \avg{v}\jump{\beta} \right) + \He_{1,4}\left( 2 \avg{\beta}\jump{w} + 2 \avg{w}\jump{\beta} \right) + \He_{1,5}\left( - 2 \jump{\beta}\right) \\
+\, \He_{1,6}\left( 2 \avg{\beta}\jump{B_1} + 2 \avg{B_1}\jump{\beta} \right) +  \He_{1,7}\left( 2 \avg{\beta}\jump{B_2} + 2 \avg{B_2}\jump{\beta} \right) +  \He_{1,8}\left( 2 \avg{\beta}\jump{B_3} + 2 \avg{B_3}\jump{\beta} \right). \label{eq:firstrow}
\end{gather*}}
From this equation, we directly obtain the entries of the first row of the discretized entropy Jacobian,
{\small
\begin{equation}\label{eq:firstrowH}
\He_{1} = \begin{bmatrix}\rholn & \rholn\avg{u} & \rholn\avg{v} & \rholn\avg{w} & \Eline & 0 & 0 & 0 \end{bmatrix},
\end{equation}}
where we introduced additional notation for compactness
\begin{equation*}
	\pln = \frac{\rholn}{2 \betaln},\quad \Eline = \frac{\pln}{\gamma-1} + \frac{1}{2} \rholn \uavg, \quad \mbox{and} \quad \uavg = 2\left(\avg{u}^2 + \avg{v}^2 + \avg{w}^2\right)-\left(\avg{u^2} + \avg{v^2} + \avg{w^2}\right).
\end{equation*}

One finds that the forthright solution of \eqref{eq:tobesolved} leads to an asymmetric, \ie not provably entropy stable, matrix $\He{}$. Hence, it is not possible to derive a symmetric matrix such that the equality $\jump{\vec{q}} = \He{} \jump{\vec{v}}$ holds exactly for all components of $\vec{q}$.
However, if special care is taken during the expansion of the total energy term, a matrix $\H{}$ that obeys the required property can be found. It guarantees equality in all but the jump in energy term where the equality reduces to an asymptotic one. The modified jump in total energy reads
\begin{equation}
	\overline{\jump{E}} = \resizebox{.8\hsize}{!}{$ %
		\left(\frac{1}{2(\gamma-1)\betaln} + \frac{1}{2}\uavg\right)\jump{\rho} - \frac{\rholn}{2(\gamma-1)}\frac{\jump{\beta}}{(\betaln)^2} + \avg{\rho}\left(\avg{u}\jump{u} + \avg{v}\jump{v} + \avg{w}\jump{w}\right) + \sum\limits_{i=1}^{3}\big( \avg{B_i}\jump{B_i}\big)
	$} \simeq \jump{E}.
\end{equation}
Using $\overline{\jump{E}}$ in place of $\jump{E}$, we can solve \eqref{eq:tobesolved} using the previously described technique where we have $(\jump{\vec{q}})_i = (\H{}\jump{\vec{v}})_i$ for $i = \{1,2,3,4,6,7,8\}$ and $(\jump{\vec{q}})_5 \simeq (\H{}\jump{\vec{v}})_5$. We get the complete dissipation matrix
\begin{equation}\label{eq:H}
\H{} = \resizebox{.89\hsize}{!}{$ %
\begin{bmatrix}
\rholn & \rholn\avg{u} & \rholn\avg{v} & \rholn\avg{w} & \Eline & 0 & 0 & 0 \\
\rholn\avg{u} & \rholn\avg{u}^2 + \pavg & \rholn\avg{u}\avg{v} & \rholn\avg{u}\avg{w} & \left(\Eline + \pavg \right) \avg{u} & 0 & 0 & 0 \\
\rholn\avg{v} & \rholn\avg{v}\avg{u} & \rholn\avg{v}^2 + \pavg & \rholn\avg{v}\avg{w} & \left(\Eline + \pavg \right) \avg{v} & 0 & 0 & 0 \\
\rholn\avg{w} & \rholn\avg{w}\avg{u} & \rholn\avg{w}\avg{v} & \rholn\avg{w}^2 + \pavg & \left(\Eline + \pavg \right) \avg{w} & 0 & 0 & 0 \\
\Eline & \left(\Eline + \pavg \right) \avg{u} & \left(\Eline + \pavg \right) \avg{v} & \left(\Eline + \pavg \right) \avg{w} & \H_{5,5} & \tau \avg{B_1} & \tau \avg{B_2} & \tau \avg{B_3} \\
0 & 0 & 0 & 0 & \tau \avg{B_1} & \tau & 0 & 0 \\
0 & 0 & 0 & 0 & \tau \avg{B_2} & 0 & \tau & 0 \\
0 & 0 & 0 & 0 & \tau \avg{B_3} & 0 & 0 & \tau \\
\end{bmatrix},
$}
\end{equation}
with
\begin{equation*}\label{eq:alotofequations}
\H_{5,5} = \frac{1}{\rholn}\Big(\frac{(\pln)^2}{\gamma-1} + {\Eline^2}\Big) + \pavg \Big(\avg{u}^2 + \avg{v}^2 + \avg{w}^2\Big) + \tau \sum_{i=1}^{3} \Big(\avg{B_i}^2\Big),\quad
\pavg = \frac{\avg{\rho}}{2\avg{\beta}},\quad \mbox{and}\quad \tau = \frac{\avg{p}}{\avg{\rho}}.
\end{equation*}
The discrete entropy Jacobian for the compressible Euler equations is
\begin{equation}\label{eq:HEuler}
\H_{\rm Euler} = \resizebox{.85\hsize}{!}{$ %
\begin{bmatrix}
\rholn & \rholn\avg{u} & \rholn\avg{v} & \rholn\avg{w} & \Eline \\
\rholn\avg{u} & \rholn\avg{u}^2 + \pavg & \rholn\avg{u}\avg{v} & \rholn\avg{u}\avg{w} & \left(\Eline + \pavg \right) \avg{u} \\
\rholn\avg{v} & \rholn\avg{v}\avg{u} & \rholn\avg{v}^2 + \pavg & \rholn\avg{v}\avg{w} & \left(\Eline + \pavg \right) \avg{v} \\
\rholn\avg{w} & \rholn\avg{w}\avg{u} & \rholn\avg{w}\avg{v} & \rholn\avg{w}^2 + \pavg & \left(\Eline + \pavg \right) \avg{w} \\[-0.3em]
\Eline & \left(\Eline + \pavg \right) \avg{u} & \left(\Eline + \pavg \right) \avg{v} & \left(\Eline + \pavg \right) \avg{w} & \frac{1}{\rholn}\Big(\frac{(\pln)^2}{\gamma-1} + {\Eline^2}\Big) + \pavg \big(\avg{u}^2 + \avg{v}^2 + \avg{w}^2\big) \\
\end{bmatrix}.
$}
\end{equation}

Clearly, the discrete entropy Jacobian matrices \eqref{eq:H} and \eqref{eq:HEuler} are symmetric. It is straightforward, albeit laborious, to verify using Sylvester's criterion that the discrete matrices are symmetric positive definite (SPD).
Due to the structure of the dissipation term \eqref{eq:dissipation_term2}, the SPD property of the new matrices guarantees that the numerical flux \eqref{eq:entropystableflux} complies with the entropy inequality \eqref{eq:EntropyInequality} discretely.

Next, we exemplarily test the equality \eqref{eq:dissipation_term1} = \eqref{eq:dissipation_term2} for a single entry of the obtained matrix for ideal MHD \eqref{eq:H}, namely the mass flux using the initial conditions \eqref{eq:ICs}.
If we use the discrete entropy Jacobian derived in this work, we find
\begin{align*}
	\H_{1} &= \begin{bmatrix}\rholn & \rholn \avg{u} & \rholn \avg{v} & \rholn \avg{w} & \frac{\pln}{\gamma-1} + \frac{1}{2} \rholn \uavg & 0 & 0 & 0\end{bmatrix} \\
	&\Rightarrow \H_{1} \cdot \jump{\vec{v}} = 0 = \jump{\rho}
\end{align*}
as expected.

If we, however, chose a naive averaging, this equality does not hold any longer. Assume that all entries of the entropy Jacobian are given by arithmetic means of the primitive variables, \eg
\begin{align*}
	\vec{H}_{1} &= \begin{bmatrix}\avg{\rho} & \avg{\rho} \avg{u} & \avg{\rho} \avg{v} & \avg{\rho} \avg{w} & \frac{\pavg}{\gamma-1}+\frac{1}{2} \avg{\rho} \avg{\vec{u}^2}\end{bmatrix}
	\quad\mbox{we find}\\
	&\Rightarrow \vec{H}_{1} \cdot \jump{\vec{v}} \approx \num{1.25e+6} \ne \jump{\rho},
\end{align*}
which will inevitably lead to negative densities for any practical CFL coefficient.

This short numerical experiment highlights the main massage of this work that even if one uses a suitable baseline entropy conserving flux, the stabilization term still has to be constructed carefully in order to obtain a stable numerical scheme.

We can use this to create a local Lax-Friedrichs (LLF) like numerical scheme
\begin{equation}\label{eq:KEPES_LLF}
	\vec{f}^\mathrm{KEPES,LLF} = \vec{f}^\mathrm{KEPEC} - \frac{1}{2} |\lambda_{\rm max}^\mathrm{local}| \H{} \jump{\vec{v}}
\end{equation}
where
\begin{equation}
	\lambda_{\rm max}^\mathrm{local} = \max(\lambda_{\mathrm{max},\mathrm{R}},\lambda_{\mathrm{max},\mathrm{L}})
\end{equation}
is the largest of the local ideal MHD wave speeds (\ie eigenvectors of the ideal MHD system)
\newcommand{\f}{\ensuremath{\mathrm{f}}}
\renewcommand{\s}{\ensuremath{\mathrm{s}}}
\renewcommand{\a}{\ensuremath{\mathrm{a}}}
\newcommand{\E}{\ensuremath{\mathrm{E}}}
\newcommand{\D}{\ensuremath{\mathrm{D}}}
\begin{align}\label{eq:eigenvalues}
\lambda_{\pm \f} &= u \pm c_\f, \qquad \lambda_{\pm \s} = u \pm c_\s, \qquad \lambda_{\pm \a} = u \pm c_\a, \qquad \lambda_{\D,\E} = u
\shortintertext{with}
\label{eq:cacfcs}
c_\a^2& = b_1^2, \quad
c_\mathrm{f,s}^2 = \frac{1}{2}\left(a^2+\|\vec{b}\|^2 \pm \sqrt{(a^2+\|\vec{b}\|^2)^2 - 4a^2 b_1^2}\right), \\
a^2 &= \gamma \, \frac{p}{\rho}, \quad \vec{b} = \frac{\vec{B}}{\sqrt{\rho}}, \quad b_\perp^2 = b_2^2 + b_3^2, \quad\mbox{and}\quad \beta_{1,2,3} = \frac{b_{1,2,3}}{b_\perp}, \notag
\end{align}
where $c_\f$ and $c_\s$ are the fast and slow magnetoacoustic wave speeds, respectively. $c_\a$ is the Alfv\'en wave speed. In \eqref{eq:cacfcs}, the plus sign corresponds to the fast magnetoacoustic speed, $c_\f$, and the minus sign corresponds to the slow magnetoacoustic speed, $c_\s$. Some eigenvalues may coincide depending on the magnetic field strength and direction. Hence, the complete set of eigenvectors is not obtained in a straightforward way \cite{Brio1988,Cargo}.

\section{A note on computational complexity}\label{scn:comp}
We are interested in the computational complexity of our proposed scheme in comparison to the IR scheme and less complicated dissipation operators. In order to quantitatively investigate the numerical complexity we prepared a benchmark program written in \texttt{FORTRAN}. We take the full numerical flux functions from \cite[(3.20)]{Derigs2016} (IR flux) and \cite[(B.3)]{Winters2016} (KEPEC flux) and compute the fluxes from random initial conditions.

\begin{table}[h]
	\begin{minipage}[t]{0.37\textwidth}
	\centering
	\begin{tabular}{cc}
		\toprule
		IR EC flux & KEPEC flux\\
		\midrule
		\num[separate-uncertainty=true]{1.553+-0.007} & \num[separate-uncertainty=true]{0.975+-0.002} \\
		\bottomrule
	\end{tabular}
	\caption{Computing the entropy conserving flux.}
	\label{tab:EC}
	\end{minipage}
	\hfill
	\begin{minipage}[t]{0.57\textwidth}
	\centering
	\begin{tabular}{ccc}
		\toprule
		non-ES scheme & naive $\mat{H}$ matrix & new $\H$ matrix\\
		\midrule
		\num[separate-uncertainty=true]{0.227+-0.002} &
		\num[separate-uncertainty=true]{5.467+-0.007} & \num[separate-uncertainty=true]{6.063+-0.010} \\
		\bottomrule
	\end{tabular}
	\caption{Computing the stabilization terms.}
	\label{tab:ES}
	\end{minipage}\vspace*{-1em}
	\caption*{Computational time in CPUs needed for ten million iterations on a single core. We disable the automatic code optimization of the compiler (\ie \texttt{ifort~-O0~test.F90}) for a fair comparison of the numerical complexity of the algorithms.}
\end{table}

As can be seen in Table~\ref{tab:EC}, the entropy conserving flux of Ismail and Roe (IR) is about \si{60}{\%} more costly than the kinetic energy preserving entropy conserving flux (KEPEC). This might be surprising on the first hand, but one has to keep in mind that the IR flux makes use of the complex $\vec{z}$ vector parametrization, introducing a considerable amount of hidden additional complexity.

A second question goes towards the computational costs of the entropy stable fluxes where the new matrix dissipation operator we found requires very specific averages. Therefore, we want to compare the computational time needed to compute the scheme given by \eqref{eq:KEPES_LLF} to three numerical schemes:
\begin{description}
	\item[non-ES scheme] This scheme is given by a simple LLF stabilization:
	\begin{equation}
		\vec{f}^\text{non-ES,LLF}_{\phantom{123456789123}} = \vec{f}^\mathrm{KEPEC} - \frac{1}{2} |\lambda_{\rm max}^\mathrm{local}| \jump{\vec{q}}. \label{eq:nonES}
	\end{equation}
	\item[ad hoc dissipation operator] We arbitrarily choose arithmetic means in the entropy Jacobian:
	\begin{equation}
		\vec{f}^\mathrm{KEPES,naive}_{\phantom{123456789123}} = \vec{f}^\mathrm{KEPEC} - \frac{1}{2} |\lambda_{\rm max}^\mathrm{local}| \mat{H} \jump{\vec{v}}. \label{eq:LLFnaive}
	\end{equation}
	This procedure is consistent with present literature since it has not been mentioned before that special care has to be taken when selecting the dissipation operator.
	\item[new dissipation operator] For this scheme, we use the entropy Jacobian derived in this work:
	\begin{equation}
		\vec{f}^\mathrm{KEPES}_{\phantom{123456789123}} = \vec{f}^\mathrm{KEPEC} - \frac{1}{2} |\lambda_{\rm max}^\mathrm{local}| \H \jump{\vec{v}}. \label{eq:LLFnew}
	\end{equation}
	Note that the scheme given by \eqref{eq:LLFnew} is identical to \eqref{eq:KEPES_LLF}.
\end{description}
The three schemes given by \eqref{eq:nonES} -- \eqref{eq:LLFnew} rely on the same baseline flux, the kinetic energy preserving entropy conservative (KEPEC) ideal MHD flux. The only difference is in the selection of the stabilization term.

As can be seen in Table~\ref{tab:ES}, the non entropy stable, very simple dissipation term is much faster than the computation involving the entropy Jacobian matrix. However, as shown in the numerical results section, both \eqref{eq:nonES} and \eqref{eq:LLFnaive} lead to a breakdown of the scheme in the first timestep. Although \eqref{eq:LLFnew} is the most expensive stabilization in this comparison, it is the only stable scheme that does not suffer from a breakdown. We find that the computation of the new $\H$ matrix involves about \si{27}{\%} higher computational costs compared to the H matrix with ad hoc averages. This only minor increase in computational costs can be attributed to the fact that, although we have to compute more complicated averages, we precalculate them once and use the stored result when filling the matrix.

Note that a significant portion of the execution time ($\approx\SI{40}{\%}$) is spend in the \texttt{FORTRAN} function \texttt{MATMUL} used to multiply the entropy Jacobian with the jump of the entropy vector. Hence, an optimized version may increase the computational efficiency considerably.

\section{Numerical tests}\label{scn:numexp}
We illustrate the increased robustness of the scheme derived in this work by considering two simple one dimensional test problems.

\subsection{Shock in fast moving magnetized background medium - breakdown of the IR scheme}
As a first robustness test, we consider a demanding problem involving a shock traveling though a fast moving and magnetized medium. This test represents typical situations in large-scale astrophysical simulations namely supernova explosions in various environments, \eg \cite{Gatto2015}. We take the first numerical experiment given in this paper \eqref{eq:ICs} and create a full numerical test. The test is computed on a uniform grid of $256$ cells with a CFL coefficient of $0.8$ and a domain size of $x\in[-1,1]$. We select uniform density, $\rho = 1$, uniform velocity of the background $\vec{u} = [10,\,0,\,0]^\intercal$ as well as a uniform background magnetic field $\vec{B} = [\num{1e-2},\,0,\,0]$. To simulate the strong explosion, we inject a pressure discontinuity with $p_\mathrm{in} = 1.0$ at $|x_\mathrm{in}| \le 0.1$. Outside of this injection region, we set $p_\mathrm{out} = \num{1e-6}$. We use periodic boundary conditions and compute the solution at $t_\mathrm{end}=\num{5e-2}$ using an ideal gas equation of state with an adiabatic index of $\gamma=5/3$.

\begin{figure}[h]
	\centering
	\includegraphics{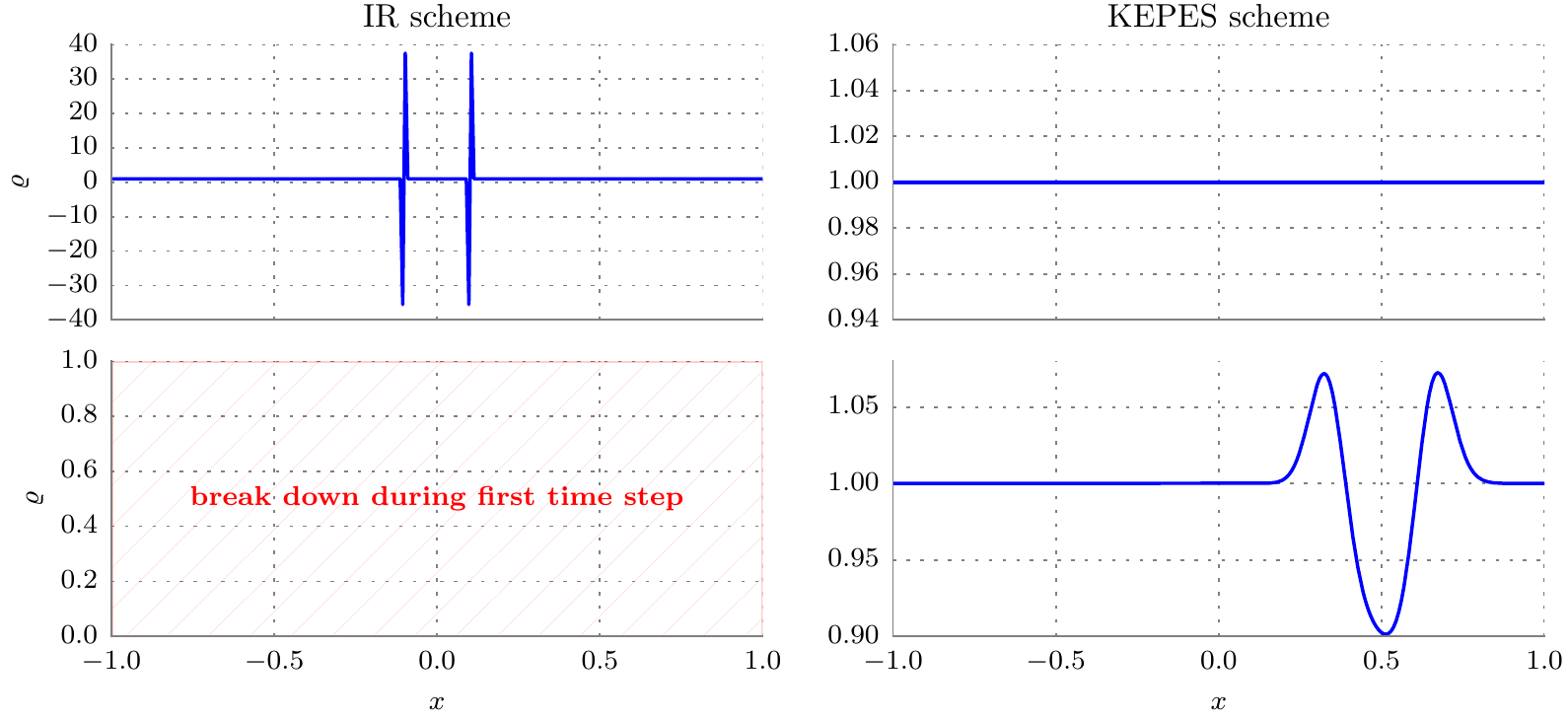}
	\caption{Result of the fast shock propagating through a magnetized medium which is moving in $x$-direction with a constant speed. We show the result of the first time step ($t\approx\num{4e-4}$, top) and the set simulation end time ($t_\mathrm{end} = \num{5e-2}$, bottom). The IR scheme (left) fails in the first time step, while the new scheme derived in this work (right) is stable.}
	\label{fig:test}
\end{figure}

The result of this test is shown in Fig.~\ref{fig:test}. We observe that the IR scheme fails during the first time step even for greatly lowered CFL coefficients. However, this test can easily be computed using the KEPEC + the new scalar dissipation flux derived in this work \eqref{eq:KEPES_LLF}. As expected, two shock fronts develop, propagating outwards. The ``explosion'' is moving to the right with the initial background velocity, leading to a final displacement of $u\cdot t_\mathrm{end} = 0.5$ at $t_\mathrm{end} = \num{5e-2}$.

\subsection{Importance of the averaging in the dissipation operator}
As a second robustness test, we use the previously described test case and use the three different stabilizations terms described in section \ref{scn:comp}. We find that the two stabilizations which are not carefully constructed for entropy stability fail for the demanding test case described above.

\begin{figure}[h]
	\centering
	\includegraphics{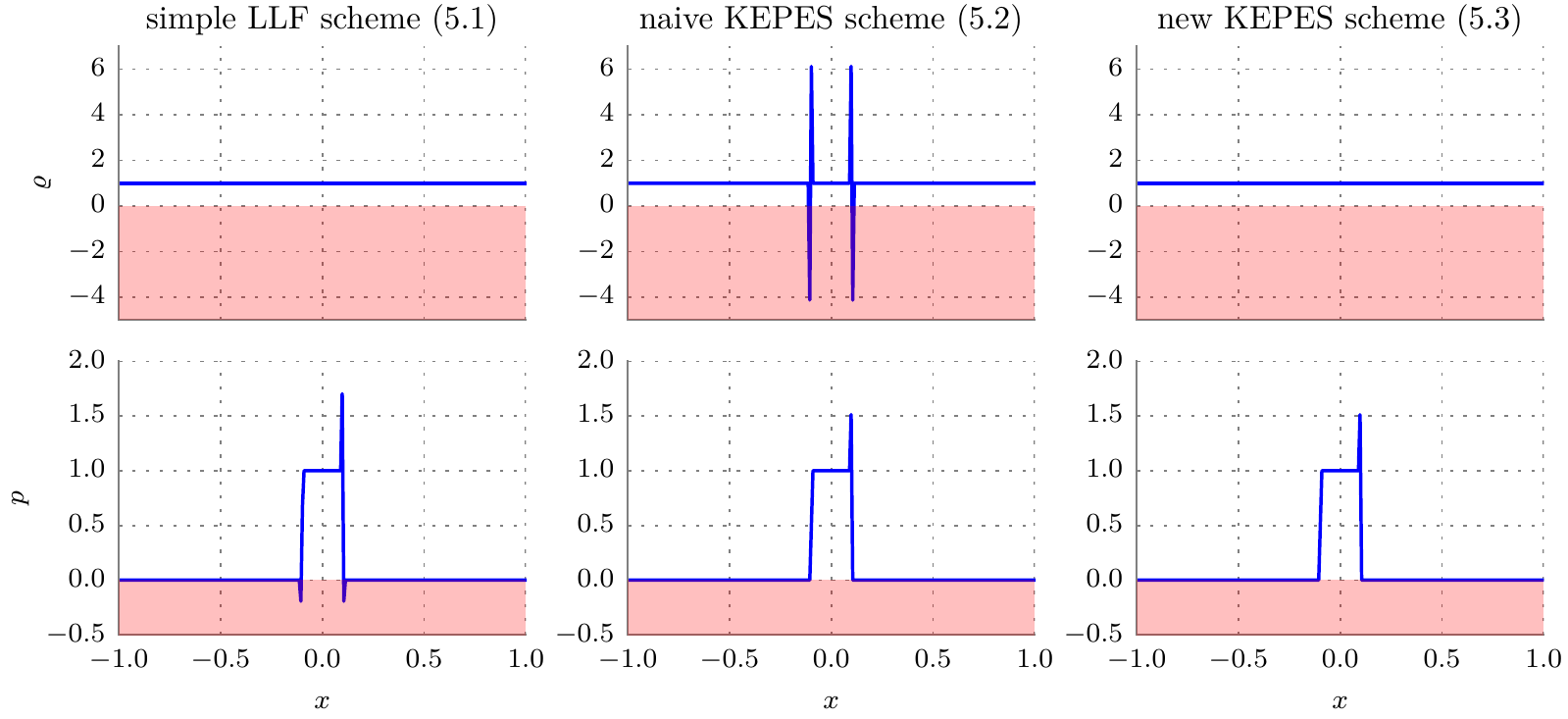}
	\caption{Density (top) and thermal pressure (bottom) plots of the shock propagating through a fast magnetized medium test at $t\approx\num{4e-4}$. The simple LLF scheme (left) over-stabilizes the total energy update, leading to a break down because of negative pressures. The KEPES scheme with the naively averaged dissipation operator (center) produces both negative densities and pressures. The scheme involving the correctly averaged dissipation operator (right) is stable.}
	\label{fig:test2}
\end{figure}

In Fig.~\ref{fig:test2} we see a failure of the simple non entropy stable LLF scheme (left) as well as of the scheme involving the naively averaged entropy Jacobian $\mat{H}$ (middle). The LLF scheme overestimates the stabilization of the total energy and hence leads to negative thermal pressures in two cells. The naive KEPES scheme breaks down because the required equalities $\jump{\vec{q}} \simeq \mat{H}\cdot\jump{\vec{v}}$ are not approximated well for very large jumps in neither the conservative nor the entropy variables. Therefore, this demanding test triggers a breakdown of the scheme, leading to large negative densities ($\min(\rho)\approx-4$) as well as small negative pressures ($\min(p)\approx\num{-1.4e-5}$). In direct comparison, we see that interchanging the discrete entropy Jacobian matrix\footnote{which is the only difference between \eqref{eq:LLFnaive} and \eqref{eq:LLFnew}} is sufficient to make the scheme robust.

\section{Conclusion}\label{scn:Conclusion}
In this work we present a technique that is convenient for the derivation of discrete entropy Jacobian operators. We exemplarily apply the technique to a very simple choice of a Rusanov-like mean value entropy Jacobian that guarantees fulfillment of the entropy inequality for ideal MHD and the compressible Euler equations. It emerges that a unique averaging technique is required and it is shown that a discrete SPD entropy Jacobian matrix can be obtained choosing specific averages during the expansion of the jump in total energy.

As a baseline flux, we choose the kinetic energy preserving entropy conservative flux recently presented in \cite{Chandrashekar2012,Winters2016} since we experience unphysical results with the entropy conserving flux of Ismail and Roe for shocks in moving media as we show using a simple test problem in the numerical results section.
Future work on the field of entropy-stable approximations can benefit from the knowledge of the required averaging technique presented in this paper. Follow up work will describe more complex matrix dissipation terms.

\section*{Acknowledgment}
DD and SW acknowledge the support of the Bonn-Cologne Graduate School for Physics and Astronomy (BCGS), which is funded through the Excellence Initiative, as well as the Sonderforschungsbereich (SFB) 956 on the ``Conditions and impact of star formation''. SW thanks the Deutsche Forschungsgemeinschaft (DFG) for funding through the SPP 1573 ``The physics of the interstellar medium''.

\section*{References}
\bibliography{mybibfile}

\begin{thebibliography}{10}
\expandafter\ifx\csname url\endcsname\relax
  \def\url#1{\texttt{#1}}\fi
\expandafter\ifx\csname urlprefix\endcsname\relax\def\urlprefix{URL }\fi
\expandafter\ifx\csname href\endcsname\relax
  \def\href#1#2{#2} \def\path#1{#1}\fi

\bibitem{Fjordholm2012}
U.~S. Fjordholm, S.~Mishra, E.~Tadmor, {Arbitrarily High-order Accurate Entropy
  Stable Essentially Nonoscillatory Schemes for Systems of Conservation Laws},
  SIAM Journal on Numerical Analysis 50~(2) (2012) 544--573.
\newblock \href {http://dx.doi.org/10.1137/110836961}
  {\path{doi:10.1137/110836961}}.

\bibitem{Derigs2016}
D.~Derigs, A.~R. Winters, G.~J. Gassner, S.~Walch, {A Novel High-Order, Entropy
  Stable, 3D AMR MHD Solver with Guaranteed Positive Pressure}, Journal of
  Computational Physics (2016) --\href
  {http://dx.doi.org/10.1016/j.jcp.2016.04.048}
  {\path{doi:10.1016/j.jcp.2016.04.048}}.

\bibitem{Winters2016}
A.~R. Winters, G.~J. Gassner, {Affordable, entropy conserving and entropy
  stable flux functions for the ideal MHD equations}, Journal of Computational
  Physics 304 (2016) 72--108.
\newblock \href {http://dx.doi.org/10.1016/j.jcp.2015.09.055}
  {\path{doi:10.1016/j.jcp.2015.09.055}}.

\bibitem{Fjordholm2011}
U.~S. Fjordholm, S.~Mishra, E.~Tadmor, {Well-Blanaced and Energy Stable Schemes
  for the Shallow Water Equations with Discontiuous Topography}, Journal of
  Computational Physics 230~(14) (2011) 5587--5609.
\newblock \href {http://dx.doi.org/10.1016/j.jcp.2011.03.042}
  {\path{doi:10.1016/j.jcp.2011.03.042}}.

\bibitem{Tadmor1987}
E.~Tadmor, The numerical viscosity of entropy stable schemes for systems of
  conservation laws, Mathematics of Computation 49~(179) (1987) 91--103.
\newblock \href {http://dx.doi.org/10.2307/2008251}
  {\path{doi:10.2307/2008251}}.

\bibitem{Barth1999}
T.~J. Barth, {Numerical Methods for Gasdynamic Systems on Unstructured Meshes},
  in: D.~Kr\"{o}ner, M.~Ohlberger, C.~Rohde (Eds.), {An Introduction to Recent
  Developments in Theory and Numerics for Conservation Laws}, Vol.~5 of Lecture
  Notes in Computational Science and Engineering, Springer Berlin Heidelberg,
  1999, pp. 195--285.
\newblock \href {http://dx.doi.org/10.1007/978-3-642-58535-7_5}
  {\path{doi:10.1007/978-3-642-58535-7_5}}.

\bibitem{Roe2006}
P.~L. Roe, {Affordable, entropy consistent flux functions}, in: {Eleventh
  International Conference on Hyperbolic Problems: Theory, Numerics and
  Applications}, Lyon, 2006.

\bibitem{IsmailRoe2009}
F.~Ismail, P.~L. Roe, {Affordable, entropy-consistent {E}uler flux functions
  {II}: Entropy production at shocks}, Journal of Computational Physics
  228~(15) (2009) 5410--5436.
\newblock \href {http://dx.doi.org/10.1016/j.jcp.2009.04.021}
  {\path{doi:10.1016/j.jcp.2009.04.021}}.

\bibitem{Chandrashekar2012}
P.~Chandrashekar, {Kinetic Energy Preserving and Entropy Stable Finite Volume
  Schemes for Compressible Euler and Navier-Stokes Equations}, Communications
  in Computational Physics 14 (2013) 1252--1286.
\newblock \href {http://dx.doi.org/10.4208/cicp.170712.010313a}
  {\path{doi:10.4208/cicp.170712.010313a}}.

\bibitem{Brio1988}
M.~Brio, C.~C. Wu, An upwind differencing scheme for the equations of ideal
  magnetohydrodynamics, Journal of Computational Physics 75~(2) (1988)
  400--422.
\newblock \href {http://dx.doi.org/10.1016/0021-9991(88)90120-9}
  {\path{doi:10.1016/0021-9991(88)90120-9}}.

\bibitem{Cargo}
P.~Cargo, G.~Gallice, {Roe Matrices for Ideal MHD and Systematic Construction
  of Roe Matrices for Systems of Conservation Laws}, Journal of Computational
  Physics 136~(2) (1997) 446 -- 466.
\newblock \href {http://dx.doi.org/10.1006/jcph.1997.5773}
  {\path{doi:10.1006/jcph.1997.5773}}.

\bibitem{Gatto2015}
A.~Gatto, S.~Walch, M.-M.~M. Low, T.~Naab, P.~Girichidis, S.~C.~O. Glover,
  R.~W\"unsch, R.~S. Klessen, P.~C. Clark, C.~Baczynski, T.~Peters, J.~P.
  Ostriker, J.~C. Ib\'a\~nez Mej\'ia, S.~Haid, {Modelling the supernova-driven
  ISM in different environments}, Monthly Notices of the Royal Astronomical
  Society 449~(1) (2015) 1057--1075.
\newblock \href {http://dx.doi.org/10.1093/mnras/stv324}
  {\path{doi:10.1093/mnras/stv324}}.

\end{thebibliography}

\appendix
\section{Application of Sylvester's criterion}
We give here the eight determinants of the leading principal minors ($M_{1-8}$) of the obtained matrix for ideal MHD $\H$ \eqref{eq:H}. Since $\rho,p,\tau > 0, \gamma > 1$, the SPD property is immediately clear.
\begin{align*}
\left|M_{1}\right| &= \left|\left[\rholn\right]\right| = \rholn,\\[0.1cm]
\left|M_{2}\right| &= \left|\begin{bmatrix}
\rholn & \rholn \avg{u} \\
\rholn \avg{u} & \rholn \avg{u}^2 + \pavg \\
\end{bmatrix}\right| = \rholn \pavg,\\[0.1cm]
\left|M_{3}\right| &= \left|\begin{bmatrix}
\rholn & \rholn\avg{u} & \rholn\avg{v}\\
\rholn\avg{u} & \rholn\avg{u}^2 + \pavg & \rholn\avg{u}\avg{v}\\
\rholn\avg{v} & \rholn\avg{v}\avg{u} & \rholn\avg{v}^2 + \pavg\\
\end{bmatrix}\right| = \rholn (\pavg)^2,\\[0.1cm]
\left|M_{4}\right| &= \dots = \rholn (\pavg)^3,\\[0.1cm]
\left|M_{5}\right| &= \dots = \rholn(\pavg)^3 \left(\tau\left(\avg{B_1}^2+\avg{B_2}^2+\avg{B_3}^2\right) + 2\frac{\pln}{\gamma-1}\frac{\pln}{\rholn}\right),\\[0.1cm]
\left|M_{6}\right| &= \dots =  \rholn (\pavg)^3 \tau\left(\tau\left(\avg{B_2}^2+\avg{B_3}^2\right) + 2\frac{\pln}{\gamma-1}\frac{\pln}{\rholn}\right),\\[0.1cm]
\left|M_{7}\right| &= \dots =  \rholn (\pavg)^3 \tau^2\left(\tau\avg{B_3}^2 + 2\frac{\pln}{\gamma-1}\frac{\pln}{\rholn}\right),\\[0.1cm]
\left|M_{8}\right| &= |\H{}| = \dots = 2\frac{(\avg{p})^3(\pln)^2\tau^3}{(\gamma-1)}.
\end{align*}
\end{document}